\documentclass[twocolumn,showpacs,showkeys,nofootinbib,aps]{revtex4}
\usepackage{pstricks,graphicx,epsfig,color,amssymb,amsmath,amscd}

\newcommand{\bi}{\bibitem}
\newcommand{\be}{\begin{eqnarray}}
\newcommand{\ee}{\end{eqnarray}}

\begin{document}

\title{Gravitational particle production in braneworld cosmology}

\author{C. Bambi}
\author{F.R. Urban}
\affiliation{Istituto Nazionale di Fisica Nucleare, 
Sezione di Ferrara, I-44100 Ferrara, Italy\\
Dipartimento di Fisica, Universit\`a degli Studi di Ferrara,
I-44100 Ferrara, Italy}

\date{\today}

\begin{abstract}
Gravitational particle production in time variable
metric of an expanding universe is efficient only when
the Hubble parameter $H$ is not too small in comparison
with the particle mass. In standard cosmology, the huge
value of the Planck mass $M_{Pl}$ makes the mechanism
phenomenologically irrelevant. On the other hand, in
braneworld cosmology the expansion rate of the early
universe can be much faster and many weakly interacting
particles can be abundantly created. Cosmological
implications are discussed.
\end{abstract}



\maketitle

{\it Introduction --}
It is a well known fact that particles can be created
by classical backgrounds such as time variable spacetime
metrics \cite{metric1, metric2} and oscillating (or any other
time--dependent) fields \cite{inflaton}.
In particular, gravitational particle production in 
time varying metric is an inevitable phenomenon which 
does not depend on particle interactions, because according 
to General Relativity all the forms of energy couple to 
gravity with the same strength (Equivalence Principle), so 
that even very weakly interacting or sterile ones can be 
abundantly created. In a cosmological environment, Robertson--Walker 
metrics are conformally flat and this implies that 
conformally coupled particles, such as massless fermions and 
vector bosons, cannot be produced \cite{parker}\footnote{However, 
quantum conformal anomaly could circumvent this exclusion principle
and allow for noticeable production of even massless gauge 
bosons \cite{dolgov}.}. On the other hand, particle masses
break conformal invariance and serves as a source of particle
creation. When the Hubble parameter $H$ is much larger than
the particle mass $m$, that is $H \gg m$, the number density
of created particle is constant \cite{metric1}
\be\label{number}
n = \frac{m^3}{24 \pi^2} \, ,
\ee
whereas for $H \ll m$ particle creation is negligible and
$n$ decreases as $1/a^3$, where $a$ is the cosmological scale
factor, due to the expansion of the universe. For the sake
of simplicity, in what follows we assume that, for $H > m$,
the particle number density is given by eq.~(\ref{number})
and particle production stops instantaneously when $H = m$.

We notice here that formula (\ref{number}) is valid for any 
power law expansion of the scale factor $a(t)\propto t^q$ 
\cite{metric1} -- in standard cosmology, for matter or 
radiation dominated universe one has $q=2/3$ or $q=1/2$, 
respectively, whereas braneworld cosmology enforces $q=1/3$ or 
$q=1/4$. However, the approximation which led to (\ref{number}) 
is found to be valid only within one order of magnitude, as 
different epochs of the universe would result in different 
number densities today \cite{full}. In particular, braneworld 
regimes tend to produce roughly 10 times more fermions than 
standard cosmologies. While we will work with the analytic 
expression (\ref{number}) throughout the paper, we will 
present more realistic results when computing numerical values.

{\it Standard cosmology --}
In standard cosmology, the phenomenon is usually negligible.
The universe expansion rate during the radiation dominated 
epoch is
\be
H_{SC} = \left(\frac{8 \pi^3 g_*}{90}\right)^{1/2} 
\frac{T^2}{M_{Pl}} \, , 
\ee
where $g_*$ is the effective number of relativistic degrees
of freedom and $T$ the universe temperature. Gravitational
production of particles with mass $m_X$ stops when $H_{SC} = m_X$
at the temperature (assuming it is below the reheating 
temperature, that is, the highest temperature at which
thermodynamical equilibrium had been established)
\be
T_{SC} \simeq 9 \cdot 10^9 
\left(\frac{m_X}{100 \; {\rm GeV}}\right)^{1/2}
\left(\frac{100}{g_*}\right)^{1/4} \; {\rm GeV} \, .
\ee

If the particle is stable, or if its lifetime is longer than
the present age of the universe, its contribution to 
the total energy of the universe today would be
\be
\Omega_X \simeq 2 \cdot 10^{-17} 
\left(\frac{m_X}{100 \; {\rm GeV}}\right)^{5/2} \, ,
\ee
where a dilution factor of $\sim 0.1$ has been taken into
account.

{\it Braneworld cosmology --}
In braneworld cosmology the picture can be much more 
interesting\footnote{For instance, gravitational particle 
production in this framework has been considered in the 
context of inflation in \cite{ed}.}.
Focusing on the case with one extra dimension compactified
on a circle, the effective four dimensional Friedman equation
is \cite{brane-c}
\be
H^2 = \frac{8 \pi \rho}{3 M_{Pl}^2}
\left(1 + \frac{\rho}{2 \Lambda}\right) \, ,
\ee
where $\rho$ is the energy density of ordinary matter on the brane,
\be
\Lambda = \frac{48 \pi M_*^6}{M_{Pl}^2}
\ee
is the brane tension and $M_*$ the true gravity scale of
the five dimensional theory. The transition temperature $T_*$
is the temperature at which the evolution of the universe
switches from braneworld regime to standard one and,
if the universe is radiation dominated, it is
\be
T_* &=& 2 \left(\frac{180}{\pi g_*}\right)^{1/4}
\left(\frac{M_*^3}{M_{Pl}}\right)^{1/2} \nonumber\\
&\simeq& 20 \left(\frac{100}{g_*}\right)^{1/4}
\left(\frac{M_*}{10^5 \; {\rm GeV}}\right)^{3/2} \; {\rm MeV} \, .
\ee

For $T > T_*$, the universe is in braneworld regime,
and the expansion rate is faster than the standard one 
\be
H_{BC} = \frac{\pi^2 g_*}{180} \, \frac{T^4}{M_*^3}\, .
\ee
The freeze-out temperature of gravitational particle 
production is
\be\label{freeze-out2}
T_{BC} &\simeq& 1.2 \cdot 10^4 
\left(\frac{100}{g_*}\right)^{1/4} \cdot \nonumber\\
&& \cdot \left(\frac{m_X}{100 \; {\rm GeV}}\right)^{1/4}
\left(\frac{M_*}{10^5 \; {\rm GeV}}\right)^{3/4} \; {\rm GeV} \, .
\ee

In order for the freeze--out temperature to be higher than $T_*$,
but smaller than the five dimensional Planck mass, as required by
consistency, the following relations between $M_*$ and $m_X$ must hold
\be\label{consistency}
0.2 \, m_X \lesssim M_* \lesssim 8 \cdot 10^{11} 
\left(\frac{m_X}{100 \; {\rm GeV}}\right)^{1/3} 
\; {\rm GeV} \, .
\ee
Hence, if these limits are satisfied, a period of braneworld
gravitational particle production may\footnote{The smallest
$T_{BC} / M_*$ ratio within these limits is about $10^{-3}$.} have
taken place and, if the particle $X$ is stable or quasi-stable,
$\Omega_X$ today would be 
\be
\Omega_X = 2 \left(\frac{m_X}{100 \; {\rm GeV}}\right)^{13/4}
\left(\frac{10^5 \; {\rm GeV}}{M_*}\right)^{9/4} \, ,
\ee
where as before a dilution factor of $\sim 0.1$ has been 
accounted for.

{\it Phenomenology --}
If the transition temperature $T_*$ is smaller than the 
reheating temperature $T_R$, i.e. the universe went 
through a period of braneworld cosmology after inflation,
gravitational particle production could have been very efficient.
As it has been shown previously, this inequality depends upon the 
five dimensional gravity mass scale, and one should verify
that the freeze--out temperature for gravitational interaction 
is indeed higher that $T_*$, see eq. (\ref{consistency}).
On the other hand, if $T_* > T_R$, when the universe 
exited the inflationary period it started expanding 
following the standard Friedman equation and the existence
of extra dimensions was essentially irrelevant, as long as towers of
KK modes do not play any relevant r\^ole. 

Let us now consider possible implications of this picture
for the contemporary universe. If there existed a stable or 
quasi--stable weakly interacting or sterile (i.e. which 
interacts only gravitationally) particle $X$, it would
contribute to the cosmological dark matter today. If we require
$\Omega_X = \Omega_{DM}$ and $\Omega_{DM}^{obs} = 0.25$, 
we find
\be
M_* & \simeq & 2 \cdot 10^5 
\left(\frac{m_X}{100 \; {\rm GeV}}\right)^{13/9} 
\; {\rm GeV} \, , \nonumber\\
T_* & \simeq & 100 
\left(\frac{m_X}{100 \; {\rm GeV}}\right)^{13/6}
\; {\rm MeV} \, .
\ee

It is noteworthy that, if such a particle were sterile, 
it would be essentially impossible to produce in collider 
experiments. Moreover, if that were the case, also the early 
universe would have had only gravitational production available,
as the other mechanisms are cut off. Had we taken the numerical 
value for $n_X$ \cite{full} we would have obtained an extra 
factor of 3 for $M_*$, and $T_*$ would have been five times 
higher. We are also neglecting possible (unknown) sources of 
entropy dilution at later times.

Interesting considerations arise if the particle $X$ is
the gravitino, the supersymmetric partner of the
graviton. If it is the Lightest Supersymmetric Particle 
(LSP) and R-parity is conserved, it is stable and a good 
dark matter candidate; of course its energy density today
must not overclose the universe. If it is not the LSP, 
it is unstable and its decay products had not to spoil the 
successful predictions of the Big Bang Nucleosynthesis (BBN) and 
overclose the universe. In standard cosmology, the gravitino is 
produced out of equilibrium after inflation by particle 
inelastic scattering, supersymmetric particle decay (the 
process is relevant only if the gravitino is the LSP) and 
possible model-dependent processes involving the inflaton or 
other scalar field (dilaton, moduli, etc.). In braneworld 
cosmology, the picture is a little different, because the 
universe expansion rate is faster and KK states can be 
excited as well, see e.g. Ref. \cite{noi}. Here gravitational 
particle production is taken into account.

Let us begin with the simplest picture, where we consider
only the gravitino 0-mode and we neglect every other production
mechanisms. If the gravitino is stable and we demand 
$\Omega_{3/2} \leq \Omega_{DM}^{obs}$, we obtain
\be\label{poor-bound}
M_* \gtrsim 2 \cdot 10^5 
\left(\frac{m_{3/2}}{100 \; {\rm GeV}}\right)^{13/9} 
\; {\rm GeV} \, .
\ee
This is not a strong bound, because in models where the
gravitino is the LSP it can be very light, even in the
eV range. Moreover, since $M_*$ is constrained to be at least
around $10^4$ GeV from BBN, gravitationally produced 
gravitinos lighter than 100 GeV would be irrelevant today.

On the other hand, if the gravitino is unstable, its decay
products can alter nuclei primordial abundances. At the
BBN, the (diluted) gravitino number density to entropy
density ratio would be
\be
Y_{3/2} = 5 \cdot 10^{-11}
\left(\frac{m_{3/2}}{100 \; {\rm GeV}}\right)^{9/4}
\left(\frac{10^5 \; {\rm GeV}}{M_*}\right)^{9/4}
\ee
and, assuming that the main decay channel is hadronic,
successful BBN requires \cite{bbn}
\be\label{allowed}
Y_{3/2}^{allowed} \leq 10^{-16}
\ee
for $m_{3/2} = 1$ TeV. This implies
\be\label{bound-unstable}
M_* \gtrsim 3 \cdot 10^8 \; {\rm GeV} \, .
\ee

It is interesting to notice that the upper bound coming from 
the thermally produced gravitinos  \cite{osamu} is close to 
the lower bound obtained here. For instance, the same
$Y_{3/2}$ as in (\ref{allowed}) gives
\be\label{label-unstable-th}
M_* \lesssim 2 \cdot 10^9 \; {\rm GeV} \, .
\ee

Notice further that using numerical estimates for the 
gravitationally produced gravitinos, as borrowed from 
\cite{full}, the lower limit will become extremely close to 
(\ref{label-unstable-th}). Unfortunately other uncertainties 
affect the estimates just outlined, namely the details of the 
inflationary epoch, whether there has been further entropy 
release at late times, and so on. It is nevertheless 
remarkable how little window is left open for $M_*$ in this 
scenario.

\begin{center}
\begin{figure}
\includegraphics[width=0.48\textwidth]{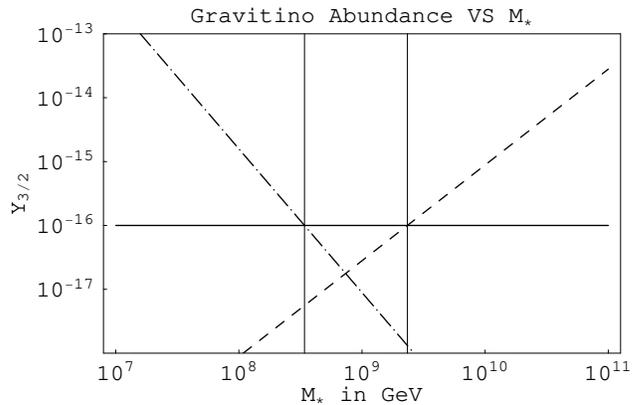}
\caption{\footnotesize{Gravitino abundances as a function 
of $M_*$, for a $m_X = 1$ TeV. Both gravitational (dot--dashed
line) and thermal (dashed line) gravitinos are 
shown. The constraint (\ref{allowed}) is the solid 
horizontal line, while the allowed range for $M_*$ in this 
case lies in the band between the two vertical lines. 
Numerical results would move the lower limit closer to the 
upper one.}}
\label{plot}
\end{figure}
\end{center}

Thus, thermally and gravitationally produced gravitinos put 
competing limits on the fundamental mass scale of the theory,
see fig. \ref{plot}.
This can be easily understood because gravitational 
production in braneworld becomes more efficient as the 
transition temperature drops (in this case the production 
stops later and there is little dilution afterwards), 
whereas the abundance of thermal gravitinos grows with it.

Of course this is the strongest limit, indeed if the
gravitino were slightly lighter or heavier, or if the
main channel were not hadronic, these constraints would 
be slightly relaxed \cite{bbn}.

As another possible application of the results obtained in
this letter, we mention the r\^ole braneworld gravitational
particle production may have in the generation of the baryon
asymmetry of the universe. This mechanism allows for
noticeable production of very weakly interacting particles
which can later on decay out of equilibrium. In this case, a
very low (TeV) gravity scale does not represent a problem, 
but is favorable for baryogenesis, as discussed in 
Ref. \cite{freese}.

{\it Conclusion --}
Particle production in time varying background metric of
an expanding universe is a well known phenomenon which is
usually only of theoretical interest in standard cosmology.
Nevertheless, in theories with extra dimensions the expansion 
rate of the early universe could have been much faster, due 
to a modified Friedman equation. This fact translates in a very 
efficient mechanism of gravitational particle production, which 
has some intriguing phenomenological implications. We have 
shown how an abundance of dark matter compatible with 
observations can be easily produced, independently on the 
specific features of the dark matter candidate, once an order 
100 GeV mass is given. This interesting feature relies on the 
fact that gravitational interactions could account for the 
necessary dark matter energy density, without the need for other 
mechanisms.

This fact holds when the LSP is the gravitino as well, provided 
that it is not too light, see eq.\ (\ref{poor-bound}). On the 
other hand, an unstable heavy gravitino is dangerous for BBN. 
Thus, safeness for BBN would strongly constrain the 
extra--dimensional scale of gravity, as equation 
(\ref{bound-unstable}) shows. Furthermore, even more interestingly, 
this lower limit is in competition with the upper limit derived 
from thermal production of gravitinos.
Once these two mechanisms are considered together there is little 
freedom in the choice of the parameters of the extra--dimensional 
model.

\begin{acknowledgments}
We wish to thank A.D. Dolgov for useful comments and encouragement, 
and A. Knauf for carefully reading the first draft of this letter.
F.U. is supported by INFN under grant n.10793/05.
\end{acknowledgments}

\end{document}